\documentstyle[12pt,eqsecnum,multicol,epsfig,aps,prb,array]{revtex}
\begin{document}
\title{Solvable models of glass transition\footnote[1]{Proceedings of the 
Cargese CNRS school {\em Physics of Glasses: structure and dynamics}, held in
Cargese, May $10-22^{th}$ 1999, edited by American Institute of Physics}}
\author{Matthieu Micoulaut \\
{\em Laboratoire de Gravitation et Cosmologie Relativistes \\ 
UPRESA 7065, Universit\'e Pierre et Marie Curie}}
\maketitle
\begin{abstract}
\par
Simple statistical agglomeration models can provide a 
universal link between the local structure and the glass transition 
temperature in network glasses.
We first stress the physical features of the models and the relevancy of the
hypothesis made and then show how to define the glass transition temperature. 
The models are applied to various types of binary, ternary and multicomponent
chalcogenide glass networks and the predictions compared to experimental data.
\par
{\bf Pacs:} 61.20N-81.20P
\end{abstract}
\section{Introduction}
Although much attention has been devoted to the understanding of the glass
transition problem\cite{Angell}, a general relationship between 
the temperature of this
transition (when measured under standard conditions, at e.g. constant heating
rate), and some easily reliable quantities is still
lacking\cite{Micoulaut}. In this paper, we show that there are some aspects of structure or 
connectivity that apparently play an important part in determining the 
absolute magnitude of $T_g$. The construction
and the prediction of this temperature from solvable agglomeration models is 
parameter-free and can be easily extended from binary to ternary, etc. glassy 
systems. 
\section{Agglomeration model}
Let us imagine a liquid that is slowly cooled and atomic motions are
progressively arrested. In network glasses, the Arrhenius-like 
increase of viscosity upon cooling to the glass transition is intimaly 
related to a decrease of dangling bonds as the starting network is polymerized. 
One typical physical
process taking place in the supercooled liquid should therefore be a kind
of sticking process in which clusters (or macromolecules) agglomerate 
together. Also, one should remark that the most important determinant of
chemical and physical properties of a glass is the concentration of 
different types of atoms involved. 
Thus the simplest level of description of such agglomeration processes
should use local structural configurations (LSC) defined by the concentration,
corresponding to short-range order (SRO), and consequently to a random network 
description of the glass. We should stress here that the next level of 
description, using intermediate range order, is very similar to the SRO 
construction \cite{IRO}. The LSC's in $Ge_xSe_{1-x}$ binary
can for instance be the germanium tetrahedrally coordinated to selenium atoms, 
or in silica based glasses the silicon tetrahedra $Q^{(k)}$ (the 
subscript k refers to the number of bonding oxygens on each $SiO_{4/2}$ 
tetrahedron). 
\par
Consider a typical cluster with a certain LSC distribution $\{p_i^0\}_{i=1..N}$. 
As long as the viscosity is not too high, other LSC can stick on this typical 
cluster, creating new covalent bonds i-j with probability: 
\begin{eqnarray}
\label{1}
p_{ij}^L(T)={\frac {W_{ij}}{\cal Z}}p_i^0p_j^0e^{-E_{ij}/k_BT}
\end{eqnarray}
where $W_{ij}$ is a statistical factor corresponding to the number of equivalent 
ways to stick a LSC $i$ on a LSC j being part of a cluster ($W_{ij}$ is 
thus related to the 
coordination numbers $m_i$ and $m_j$ of the LSC) and $E_{ij}$ is the
i-j LSC bond energy . ${\cal Z}$ normalizes the bond distribution.
The creation of these new bonds produces a local variation in the 
probability (or concentration) distribution of the cluster
and it can be encoded in the following master equation:
\begin{eqnarray}
\label{3}
{\frac {dp_i}{dt}}={\frac {1}{\tau}}\biggl[{\frac {1}{2}}\sum_{j=1}^N(1+
\delta_{ij})p_{ij}^L(T)-p_i^0\biggr]
\end{eqnarray}
where $\tau$ represents the mean agglomeration time and (\ref{3}) represents
a system of $(N-1)$ non linear differential equations. At solidification 
temperature $T_s$ ($T_s=T_m$ for a crystal or $T_s=T_g$ for a glass) one 
should reach a stationary state
and $dp_i/dt=0$, i.e. the variation of local probability distribution should be
minimized.\par
How can we distinguish a glass from a crystal ?
Imagine that a local fluctuation $\epsilon_i$ appears in the vicinity of a stationary
solution, satisfying a linearized version of equ. (\ref{3}). As usual, we can 
distinguish three types of singular points by means of the linearization. If all
the roots of the characteristic equation of the linearized system have a negative
real part, the solution is a stable attractor, i.e. it will show the preferential
agglomeration process, which corresponds to nucleation of the crystal. 
Thus there will be no possibility for a fluctuation to grow, and we identify 
the stable stationary solution with a crystal, and $T_s=T_m$. If all the real 
parts are positive, one gets an unstable stationary solution. If both are
present, a saddle point solution is obtained (fig. 1). The glass correponds to the 
latter characteristic, because it is neither a stable nor an unstable system, 
it has metastable character, and $T_s=T_g$. There is indeed still a chance 
for the system to escape (in other words for a fluctuation to grow) from the 
stationary saddle point and to fall on the stable crystalline attractor, which
always happen experimentally when a glass is annealed.
\par
One can extend such a description to binary and ternary glasses as well. 
First, we consider the 
case when $N=2$, i.e. when there are only two different 
types of LSC. We denote them by $A$ and $B$ with their respective 
coordination numbers $m_A$ and $m_B$. The system (\ref{3}) reduces then to
a single equation with only one variable, $p^0=x$, e.g. the probability of 
occurence of the LSC B (and set equal to the concentration of B species). 
The solution of (\ref{3}) yields:
\begin{eqnarray}
\label{binar1}
T_g={\frac {\Delta_B}{k_B\ln\biggl[{\frac {m_B(2x-1)}{m_A(x-1)}}\biggr]}}=
{\frac {T_0\ln\biggl[{\frac {m_B}{m_A}}\biggr]}{\ln\biggl[{\frac 
{m_B(2x-1)}{m_A(x-1)}}\biggr]}}
\end{eqnarray}
where $\Delta_B=E_{AB}-E_{AA}$ is introduced when computing the probabilities
$p_{AA}$ and $p_{AB}$ from equ. (\ref{1}) (we have neglected the possibility of
BB bonds because we are dealing in the following with low modified glasses
only, and below the stoichiometric composition). One can see in the second
part of equation (\ref{binar1}) that the relationship can 
be made parameter-free, by considering the limit $x\simeq 0$, when $T_g\simeq 
T_0$. The initial network (with glass transition temperature $T_0$) is made 
of A LSC only (e.g. the selenium network in $Ge_xSe_{1-x}$
systems) and one gets from the first part of equ. (\ref{binar1}): 
$\Delta_B=k_BT_0\ln [m_B/m_A]$.
In order to predict the glass transition temperature in a binary glass 
$A_{1-x}B_x$, there is just need of the coordination number of the 
involved LSC's and the initial glass transition temperature $T_0$ of the
A network. Finally, we can obtain a linear equation at the 
very beginning of structural modification:
\begin{eqnarray}
\label{binar2}
\biggl[{\frac {dT_g}{dx}}\biggr]_{x=0,T_g=T_0}={\frac {T_0}{\ln \biggl[{\frac
{m_B}{m_A}}\biggr]}}
\end{eqnarray}
\par
The application to ternary glass networks $A_{1-x-y}B_xC_y$ is slightly 
different, because when $N=3$, there are two non-linear equations to solve 
in terms of two probabilities $p_B^0=x$ and $p_C^0=y$. However, one obtains 
a saddle point solution from (\ref{3}), yielding again a parameter-free 
relationship between $x$, $y$ and $T_g$, because the new bond energy differences
$\Delta_C=E_{AC}-E_{AA}=k_BT_0\ln [m_C/mA]$ and $E_{BC}-E_{AA}=k_BT_0
\ln[m_Cm_B/m_A^2]$ are determined again from boundary conditions (from the binary 
AC glass for the former, similarly to $\Delta_B$, from the binary slope 
equation for the latter) \cite{preprint}. One
interesting quantity in such systems (and in multicomponent chalcogenides) is
the average coordination number, defined by $\bar r=m_A(1-x-y)+m_Bx+m_Cy$
(and $m_A=2$). From the saddle point solution of equ. (\ref{3}), we obtain a 
relationship between $\bar r$ and $T_g$, to be compared with experiment:
\begin{eqnarray}
\label{binar3}
\bar r= {\frac {2m_Bm_C\biggl[m_Bm_C\alpha\gamma(\gamma\alpha-\gamma-\alpha)+
2r_C\alpha^2(1-\gamma)+2r_B\gamma^2(1-\alpha)\biggr]}{\biggl(2r_C\alpha-2r_B\gamma
-r_Br_C\alpha\gamma\biggr)^2+8r_Br_C\alpha\gamma}}
\end{eqnarray}
where $\alpha=(2/m_C)^{(T_0/T_g)}$ and $\gamma=(2/m_B)^{(T_0/T_g)}$.
The slope in the limit $\bar r=2$ (i.e. x=y=0) has also a simple
expression:
\begin{eqnarray}
\label{binar4}
\biggl[{\frac {dT_g}{d\bar r}}\biggr]_{\bar r=2, T_g=T_0}={\frac {T_0}
{(m_B-2)\ln\biggl[{\frac {m_B}{2}}\biggr]+(m_C-2)\ln \biggl[{\frac 
{m_C}{2}}\biggr]}}
\end{eqnarray}
\section{Comparison with experimental data}
The obtained relationships (\ref{binar2}) and (\ref{binar3}) can be 
compared to the experimentally measured glass transition temperatures in 
binary, ternary and multicomponent chalcogenide glass systems. 
\par 
Given the initial glass transition
temperature $T_0$ of vitreous sulphur ($245~K$\cite{soufre}), selenium 
($316~K$\cite{soufre}) and tellurium ($343~K$ extrapolated from the data in 
\cite{tellurium}), we have plotted the equations (\ref{binar2}) and
({\ref{binar3}) for chalcogenides including elements of Group IV and V.
We can see that equation (\ref{binar2}) predicts the $T_g(x)$ trend at low
concentration for all the binary systems IV-VI and V-VI systems displayed
(fig.2). From obvious structural considerations, we can insert in equ. (\ref{binar2}) 
the value $m_B=4$ (Group IV) or $m_B=5$ (Group V), $r_A=2$ and $T_0$, to be 
compared with the plotted experimental measurements on glass transition 
temperatures. 
The prediction gives also an indirect evidence of the stiffness transition
(occuring at $\bar r=2.4$ following the theory of Phillips and Thorpe 
\cite{Phillips}). For $\bar r>2.4$ ($x>0.2$ in IV-VI glasses), the network 
looses its random character and chemical ordering occurs, due to the 
chemical stability composition at x=0.333. Thus the description
in terms of a random network of $A-A$ and $A-B$ bonds should fail at this
concentration. This is clearly seen for the $Ge_xSe_{1-x}$ data, which start 
to deviate from the equation (\ref{binar2}) at $x=0.18$, and even more at 
$x=0.24$, consistently with M\"ossbauer spectroscopy \cite{Bool}. The sulfide 
system behaves very similarly, as seen on fig. 2
(and still $r_B=4$), although the structure of the initial glass (x=0) is
rather different (chains and $S_8$ rings). The addition of germanium leads 
to a random network composed of $GeS_{4/2}$ linkages between S chains and 
rings. Note that there is no deviation for the Ge-Te compound at x=0.2. 
This can be 
related to the fact that $c-GeTe_2$ does not exist (in contrast with the
existence of $c-SiSe_2$, $c-GeSe_2$, etc.) and probably that chemical ordering 
probably does not occur at the same concentration. As a consequence, the 
network of $Ge_xTe_{1-x}$ can be thought as random. 
\par
Agreement of the prediction with experimental measurements is also 
obtained in V-VI network glasses. 
For all the systems displayed in figure 2, the slope equation
(\ref{binar2}) gives the correct trend in the variation of the glass transition 
temperature with network modification. 
Deviation of the stochastic prediction of (\ref{binar2}) is here also supposed 
to occur at $\bar r=2.4$, corresponding to $x=0.4$ in the As-Se system. 
There is some evidence that the Bismuth atom could be
five-fold coordinated in binary Bi-Se glasses \cite{data}. If this is the case,
the stiffness transition may occur $x\simeq 0.13$ (corresponding to $\bar r=
2.4$ and may be observable in the
$T_g$ data, as well as the deviation of the stochastic prediction of
equ. (\ref{binar2})). 
We can observe the same kind of agreement in ternary chalcogenides, where 
the glass transition temperature is given as a function of the average 
coordination number by equ. (\ref{binar3}) or (\ref{binar4}). The figure 3
shows that in very different systems (germanium chalcohalide glasses, thus
$m_B=4$, $m_C=1$, and telluride glasses) the network is
random for $\bar r<2.4$ and the description in terms of A-A, A-B, A-C and
B-C bonds only is accurate. The same deviation is observable at $\bar r=2.4$ 
which can again be interpreted by the occurence of chemical ordering due to
the stiffness transition (e.g.
occurence of a $Sb_2Te_3$ phase in the Ge-Sb-Te system). However,
we should point out that experimental measurements in the low modified
region of telluride systems (lower plot fig. 3) should be 
realized in order to definitely confirm the prediction. Other systems exhibit
the same universal trends in the glass transition temperature variation and 
the 
construction can be extended to quaternary and multicomponent chalcogenide
systems in a simple fashion \cite{preprint}.
\par
{\bf Acknowledgements}: The author would like to thank P. Boolchand, R. 
Kerner, G.G. Naumis, J.C. Phillips and M.F. Thorpe for very interesting 
comments and discussions on this subject.
\par

\newpage
\begin{figure}
\epsfig{figure=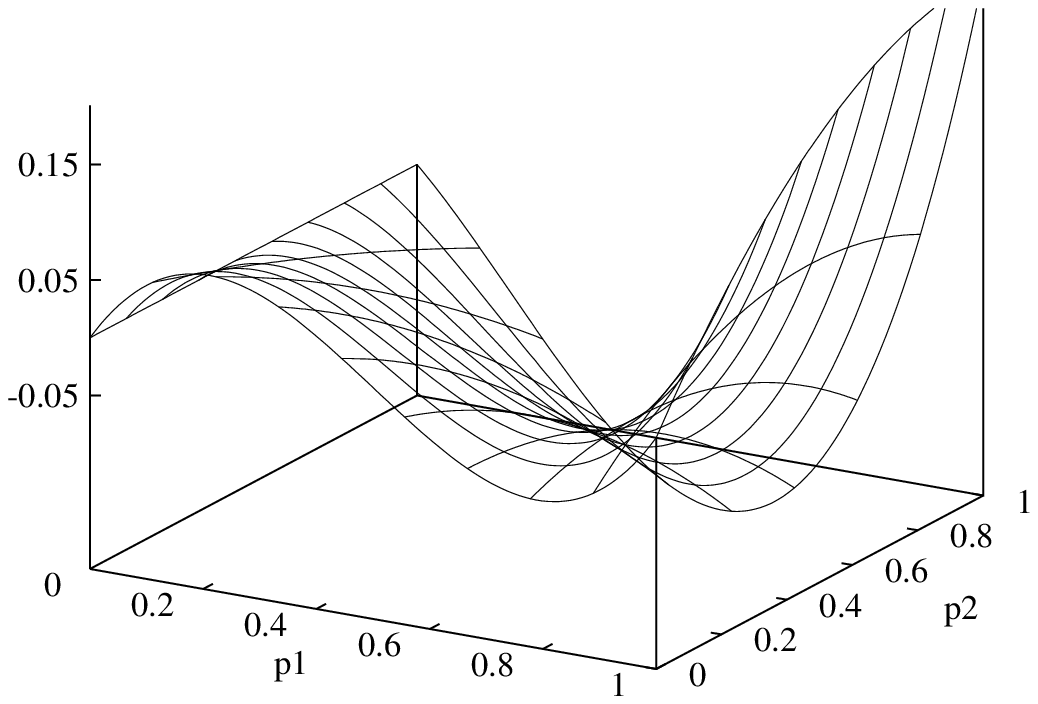} 
\caption{The right-hand side of one of the equation (\ref{3}) for $N=3$ in a simple
polygon model\cite{Dina}. The system out of equilibrium can fall on the 
attractive $p_3=1$ stationary solution (crystallization, and 
$T_s=T_m$) and never on $p_2=1$ or $p_1=1$. In 
some situations, the liquid can stay in the metastable state characterized 
by the saddle point solution at ($p_1=0.6$, $p_2=0.2$, $p_3=0.2$) and 
$T_s=T_g$. Note that the plot has to be truncated in order to have
$p_1+p_2+p_3=1$.}
\end{figure}
\newpage
\begin{figure}
\epsfig{figure=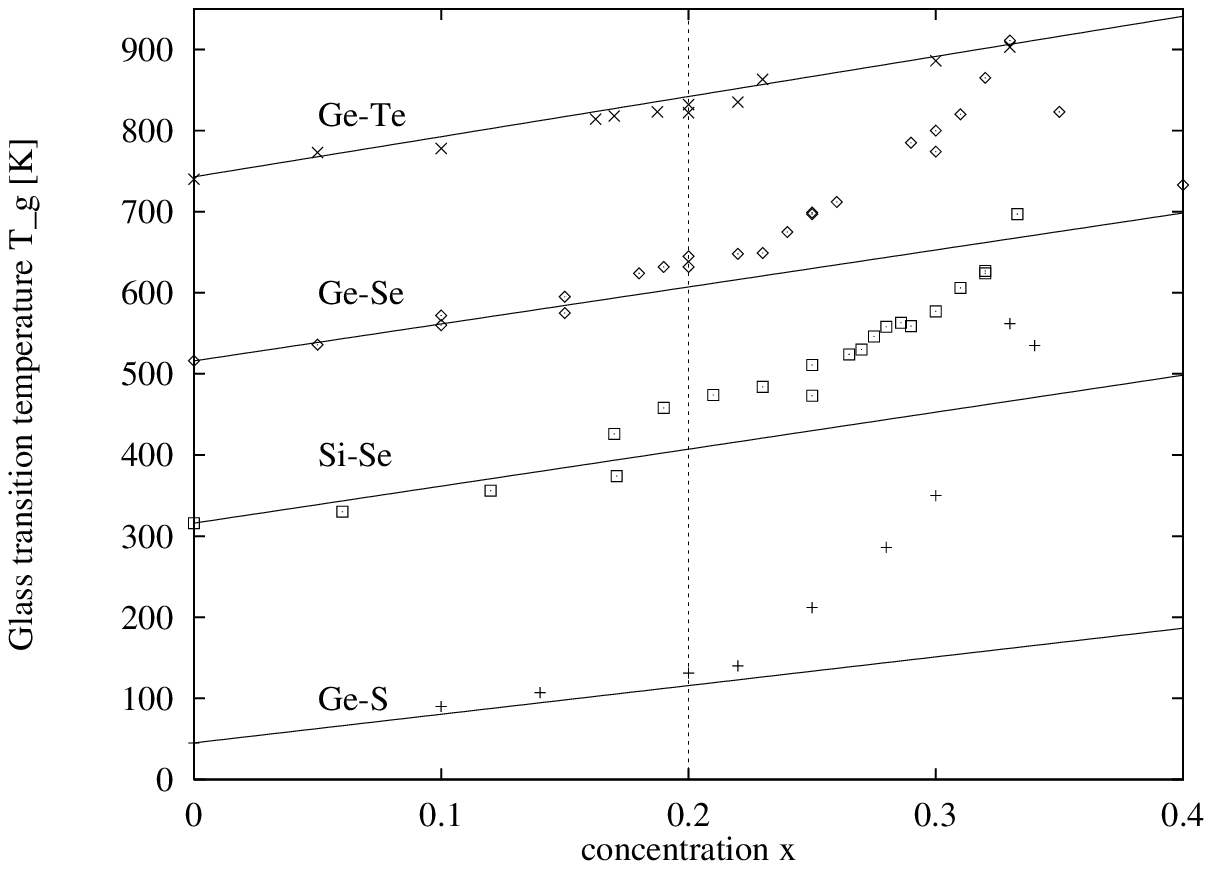}  
\vspace{1cm}
\epsfig{figure=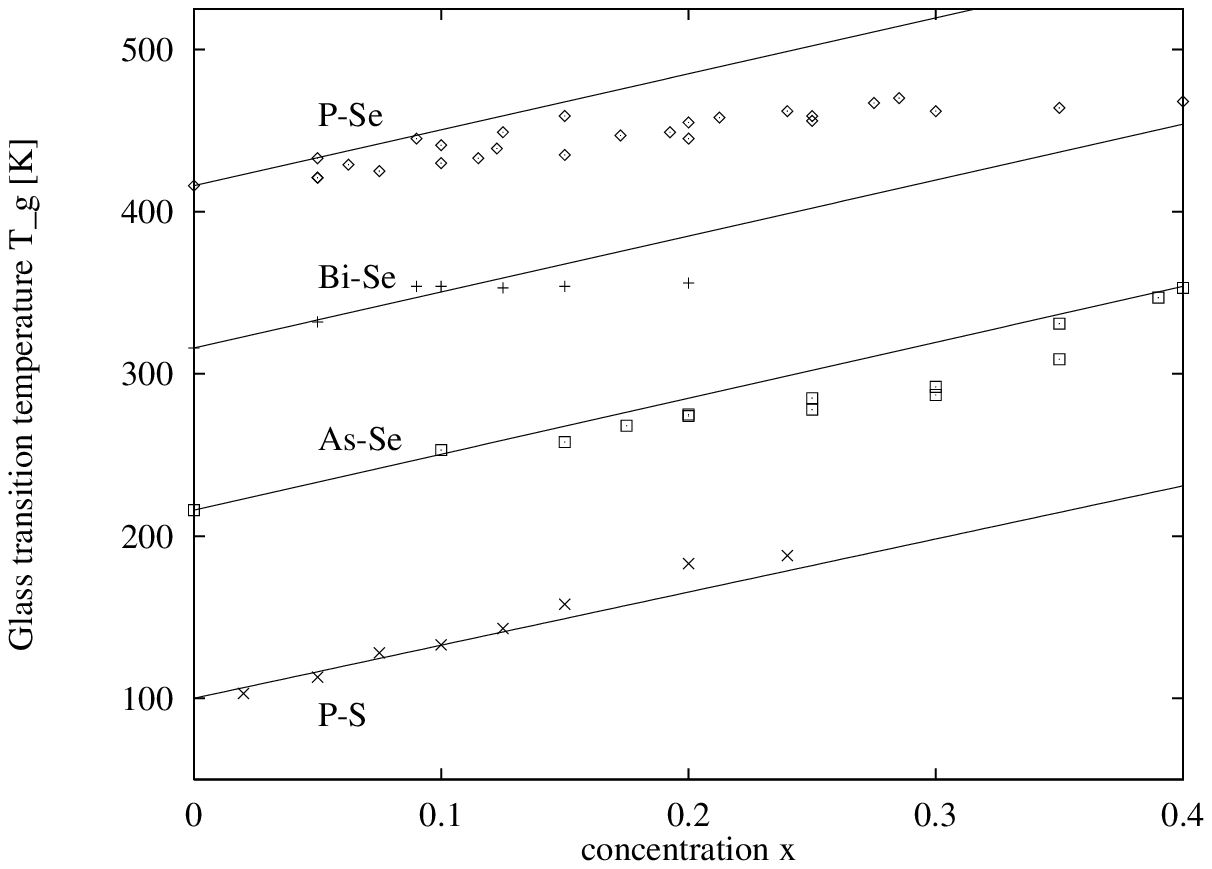} 
\vspace{0.2cm}
\caption{Binary IV-VI (upper plot) and V-VI (
lower plot) chalcogenide glasses (e.g. $Ge_xSe_{1-x}$). The lines represent the 
slope equation (\ref{binar2}) with 
$m_A=2$ and $m_B=4$ for the IV-VI glasses, and $m_B=5$ for V-VI glasses. 
Data have been displaced by 200 K and 100
K for a clearer presentation\cite{data}. The vertical shaded line corresponds 
to the critical average coordination number $\bar r_c=2.4$, predicted by
Phillips and Thorpe\cite{Phillips}.}
\end{figure}
\newpage
\begin{figure}
\epsfig{figure=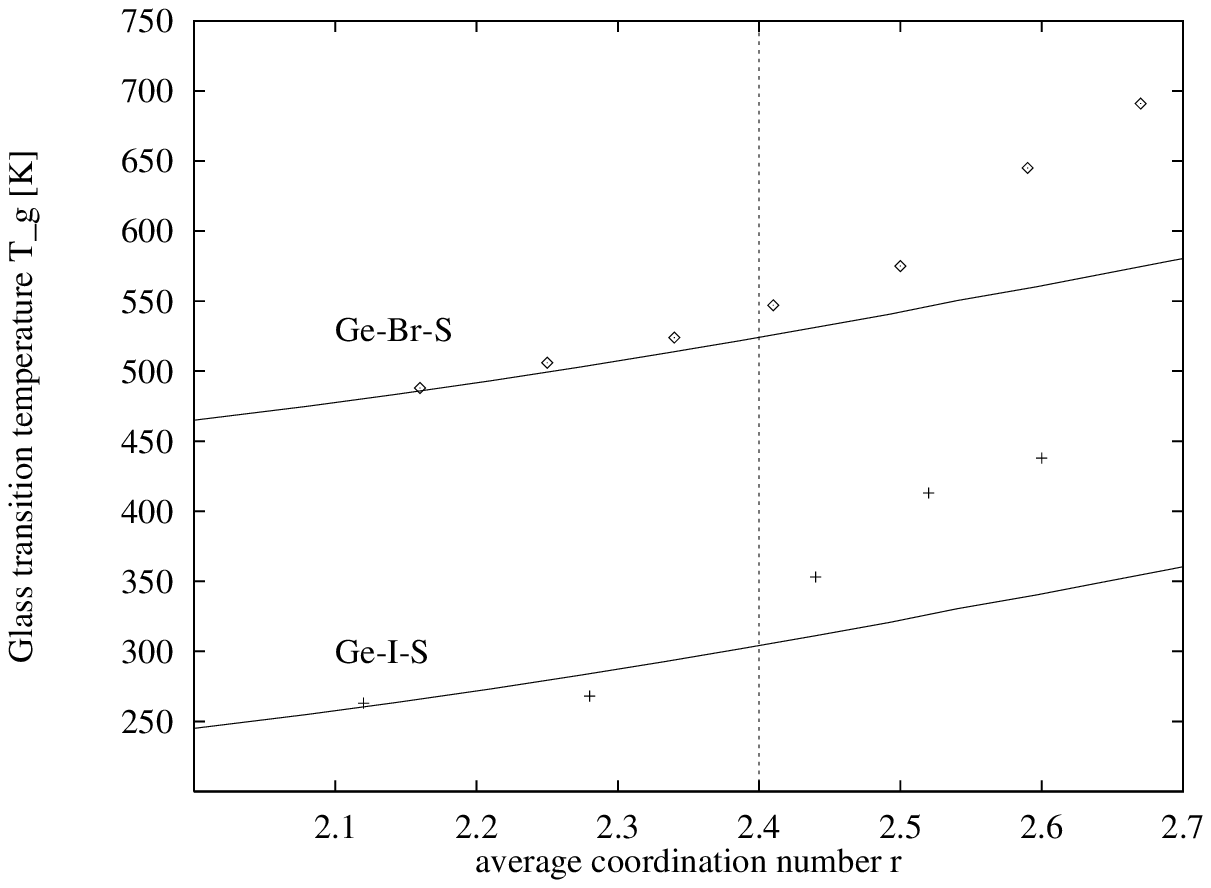}  
\vspace{2cm}
\epsfig{figure=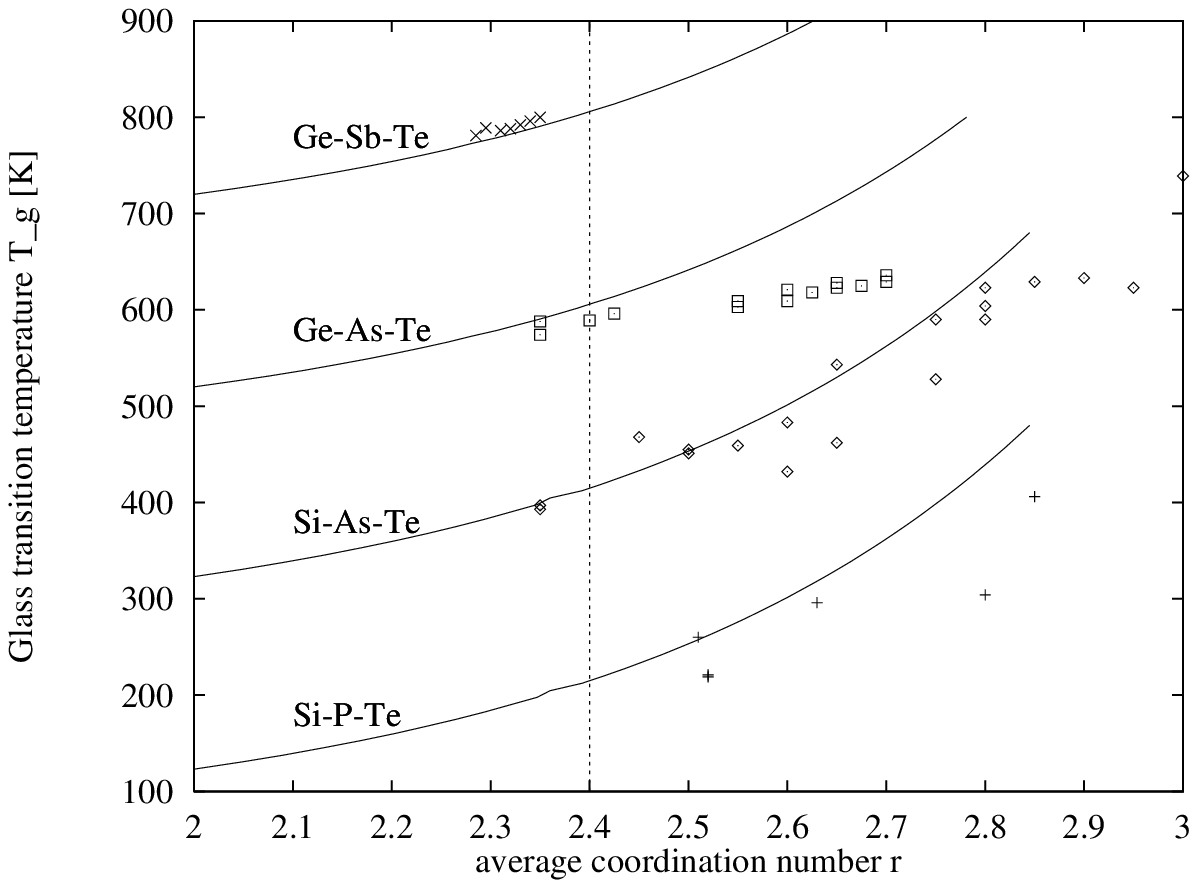}
\vspace{0.2cm}
\caption{\footnotesize Glass transition temperature in
ternary chalcogenides as a function of average coordination number $\bar r$. 
Upper plot: chalcohalide glasses. Lower plot: ternary tellurides. The curves represent 
equ. (\ref{binar3}) with $m_A=2$ and $m_B$ and $m_C$ 
inserted following the Group of the Periodic Table. Data have been displaced
for simplicity and are taken from \cite{ternar}}
\end{figure}

\begin{thebibliography}{100}
\bibitem{Angell} read this book; C.A. Angell, Science {\bf 267} (1995) 1924; 
\bibitem{Micoulaut} For a review of all previous relationships (empirical
or heuristic), see M. Micoulaut, Eur. Phys. J. B {\bf 1} (1998) 277
\bibitem{IRO} M. Micoulaut and R. Kerner, J. Phys: Cond. Matt. {\bf 9} (1997)
2551
\bibitem{Dina} R. Kerner and D.M. Dos Santos, Phys. Rev. B{\bf 37} (1988) 3881
\bibitem{preprint} M. Micoulaut, R. Kerner and G.G. Naumis, in preparation
\bibitem{soufre} S.R. Elliott, Physics of Amorphous Materials, Wiley 1989
\bibitem{tellurium} D.J. Sarrach and J.P. Deneufville, J. Non-Cryst. Solids
{\bf 22} (1976 245
\bibitem{data} G. Saffarini, Appl. Phys. A{\bf 59} (1994) 385; D. Selvanathan
MS Thesis University Cincinnati; X. Feng, W. Bresser and P. Boolchand, Phys.
Rev. Lett. {\bf 78} (1997) 4422; A. Feltz, H. Aust and D. Blayer, J. Non-Cryst.
Solids {\bf 55} (1983) 179; Y. Monteil and H. Vincent, Z. Anorg. Allg. Chem.
{\bf 428} (1977) 259; D. Lathrop, M. Tullius, T. Tepe and H. Eckert, J. 
Non-Cryst. Solids {\bf 128} (1991) 208; M.B. Myers, J.C. Schotmiller and
W.J. Hillegas, Anal. Calor. {\bf 2} (1970) 309
\bibitem{Phillips} J.C. Phillips, J. Non-Cryst. Solids {\bf 34} (1979) 153;
M.F. Thorpe, J. Non-Cryst. Solids {\bf 57} (1983) 355
\bibitem{Bool} W. Bresser, P. Boolchand and P. Suranyi, Phys. Rev. Lett.
{\bf 56} (1986) 2493
\bibitem{ternar} J. Heo and J.D. Mackenzie, J. Non-Cryst. Solids {\bf 111}
(1989) 29; A.B. Seddon and M.A. Hemingway, J. Non-Cryst. Solids {\bf 161}
(1993) 323; A. Srinivasan and E.S.R. Gopal, J. mater. Sci. {\bf 27} (1992)
4208; P. Lebaudy, J.M. Saiter, J. Grenet, M. Belhadji and C. Vautier, J. Mater.
Sci. Eng. {\bf 132A} (1991) 273
\end{thebibliography}
\end{document}